\documentclass[dvips]{acta}
\usepackage{supertabular,lscape,epsfig}
\usepackage{amssymb}
\usepackage{amsmath}
\DeclareSymbolFont{ppa}{OT1}{ppl}{m}{it}
\DeclareMathSymbol{\vv}{\mathalpha}{ppa}{'166}

\newfont{\hb}{rphvb at 10pt}
\newfont{\hbo}{rphvbo at 10pt}
\newfont{\bitt}{rptmbi at 12pt}
\newfont{\bits}{rptmbi at 11pt}

\SetPages{123}{138}

\SetVol{61}{2011}

\begin{document}

\newcommand{\TabApp}[2]{\begin{center}\parbox[t]{#1}{\centerline{
  {\bf Appendix}}
  \vskip2mm
  \centerline{\small {\spaceskip 2pt plus 1pt minus 1pt T a b l e}
  \refstepcounter{table}\thetable}
  \vskip2mm
  \centerline{\footnotesize #2}}
  \vskip3mm
\end{center}}

\newcommand{\TabCapp}[2]{\begin{center}\parbox[t]{#1}{\centerline{
  \small {\spaceskip 2pt plus 1pt minus 1pt T a b l e}
  \refstepcounter{table}\thetable}
  \vskip2mm
  \centerline{\footnotesize #2}}
  \vskip3mm
\end{center}}

\newcommand{\TTabCap}[3]{\begin{center}\parbox[t]{#1}{\centerline{
  \small {\spaceskip 2pt plus 1pt minus 1pt T a b l e}
  \refstepcounter{table}\thetable}
  \vskip2mm
  \centerline{\footnotesize #2}
  \centerline{\footnotesize #3}}
  \vskip1mm
\end{center}}

\newcommand{\MakeTableApp}[4]{\begin{table}[p]\TabApp{#2}{#3}
  \begin{center} \TableFont \begin{tabular}{#1} #4 
  \end{tabular}\end{center}\end{table}}

\newcommand{\MakeTableSepp}[4]{\begin{table}[p]\TabCapp{#2}{#3}
  \begin{center} \TableFont \begin{tabular}{#1} #4 
  \end{tabular}\end{center}\end{table}}

\newcommand{\MakeTableee}[4]{\begin{table}[htb]\TabCapp{#2}{#3}
  \begin{center} \TableFont \begin{tabular}{#1} #4
  \end{tabular}\end{center}\end{table}}

\newcommand{\MakeTablee}[5]{\begin{table}[htb]\TTabCap{#2}{#3}{#4}
  \begin{center} \TableFont \begin{tabular}{#1} #5 
  \end{tabular}\end{center}\end{table}}

\newfont{\bb}{ptmbi8t at 12pt}
\newfont{\bbb}{cmbxti10}
\newfont{\bbbb}{cmbxti10 at 9pt}
\newcommand{\uprule}{\rule{0pt}{2.5ex}}
\newcommand{\douprule}{\rule[-2ex]{0pt}{4.5ex}}
\newcommand{\dorule}{\rule[-2ex]{0pt}{2ex}}
\def\thefootnote{\fnsymbol{footnote}}
\begin{Titlepage}
\Title{The Optical Gravitational Lensing Experiment.\\
Dwarf Novae in the OGLE Data.\\ I. Three New Dwarf Novae: One in the Period
Gap and Two Longer Period Objects\footnote{Based on observations obtained
with the 1.3~m Warsaw telescope at the Las Campanas Observatory of the
Carnegie Institution for Science.}}

\Author{R.~~P~o~l~e~s~k~i$^1$,~~ 
A.~~U~d~a~l~s~k~i$^1$,~~ 
J.~~S~k~o~w~r~o~n$^2$,~~ 
M.~K.~~S~z~y~m~a~ñ~s~k~i$^1$,\\
M.~~K~u~b~i~a~k$^1$,~~
G.~~P~i~e~t~r~z~y~ñ~s~k~i$^{1,3}$,~~ 
I.~~S~o~s~z~y~ñ~s~k~i$^1$,~~
S.~~K~o~z~³~o~w~s~k~i$^1$,\\ 
P.~~P~i~e~t~r~u~k~o~w~i~c~z$^1$,~~
£.~~W~y~r~z~y~k~o~w~s~k~i$^{1,4}$~~and~~K.~~U~l~a~c~z~y~k$^1$}
{$^1$Warsaw University Observatory, Al. Ujazdowskie 4, 00-478 Warszawa, Poland\\
e-mail:
(rpoleski,udalski,msz,mk,pietrzyn,soszynsk,simkoz,pietruk,wyrzykow,kulaczyk)\\
@astrouw.edu.pl\\
$^2$Department of Astronomy, Ohio State University, 140~W. 18th~Ave.,
Columbus, OH~43210, USA
\\ e-mail: jskowron@astronomy.ohio-state.edu\\ 
$^3$Universidad de Concepción, Departamento de Astronomia, Casilla~160-C,
Concepción, Chile\\ 
$^4$~Institute of Astronomy, University of Cambridge, Madingley Road, Cambridge CB3~0HA,~UK}
\Received{June 29, 2011}
\end{Titlepage}

\Abstract{
We report serendipitous discovery of three new dwarf novae which eruptions
in 2010 were observed by the ongoing microlensing survey OGLE-IV. All three
objects are located in the Galactic bulge fields observed with the highest
cadence of 20 minutes. In the OGLE-III and OGLE-IV data we revealed a total
of 23 outbursts for one of the stars. What makes this object most
interesting is the derived superhump period of 2.61~h placing it in the
orbital period gap. The superhump period changed during the superoutburst
with a very short timescale. For two other objects, for which we observed
outburst, the orbital periods of 5.4~h and 9.5~h were measured in the
quiescence.}{novae, cataclysmic variables -- binaries: close -- surveys}

\Section{Introduction}
Because outbursts of dwarf novae (DNe) are unpredictable astrophysical
phenomena, discovering new objects of this class has been, for the very
long time, domain of amateur astronomers. The situation has changed in the
last couple of years when massive sky surveys started regular monitoring of
large areas of the sky or the densest stellar regions in the sky. Tens new
cataclysmic variables (CVs) were detected by the SDSS sky survey (\eg
Szkody \etal 2007, Southworth \etal 2010) or microlensing surveys
(Cieslinski \etal 2003, 2004).

The OGLE (Optical Gravitational Lensing Experiment) microlensing survey is
regularly monitoring millions of stars for microlensing events toward the
Galactic center and Magellanic Clouds since early 1990s. It regularly
detected many outbursts of DNe in the observed fields. The algorithm of
real time detection system of microlensing events -- EWS system (Udalski
2003) -- in the natural way detected also brightenings of outbursting DNe
(Skowron \etal 2009). However, these objects were usually treated as the
background noise in the microlensing detection process and then
neglected. The only study of CVs based on the limited OGLE-II data was done
by Cieslinski \etal (2003).

The OGLE microlensing data collected during the first three phases of the
project OGLE-I -- OGLE-III (1992--2009) while providing very precise long
term photometry have only limited applications in the characterization of
DNe. They could be used for the detection of new DNe, constraining their
frequency of occurrence or analysis of frequency of outbursts. However, the
typical cadence of observations in these phases, namely one/two
observations per night, prevented more accurate characterization of
outbursts lasting typically a few days. On the other hand the OGLE data are
well suited and were successfully used for characterization of slower
variability CVs like, for example, Nova Sco~2008 -- V1309~Sco (Tylenda
\etal 2011).

New generation phase of the OGLE survey, OGLE-IV, that started in March
2010, provides new observing capabilities that completely change the above
limitations. Several 1.4 square degrees fields in the direction of the
Galactic bulge are now regularly observed with the cadence of 20 or 60
minutes. This enables studies of all ultra-short time variable objects
including the outbursts of even so short periodic variables as from the
SU~UMa type subgroup of DNe.

Here, in this pilot paper, we present three examples of OGLE-IV
observations of outbursts of DNe to show the potential and capabilities of
the new OGLE-IV survey. Encouraged by these early results we decided to
carry out more comprehensive search for CVs in the already collected
OGLE-IV data as well as to save all the CVs detected in real time by the
OGLE-IV EWS microlensing system and then analyze properties of the detected
objects. Results will be presented in the subsequent papers of this
series.

Dwarf novae are semidetached binary systems with Roche lobe filling main
sequence secondary which matter is accreted {\it via} a disk onto the white
dwarf primary. The most easily noticeable features in the distribution of
known DNe orbital periods (Knigge 2006) are the presence of a sharp period
minimum (76 minutes) and the period gap between 2 and 3 hours where the
number of systems found is very low. In objects with periods longer than
the upper limit of the period gap the accretion is driven by the angular
momentum loss of the secondary by magnetic braking. This process stops when
the secondary becomes fully convective which is predicted at an orbital
period of $P_{\rm orb}=2.85$~h by models of Howell \etal (2001). The
boundary value derived from the histogram of observed systems by Knigge
(2006) is 3.18~h with smaller discontinuity seen close to the value of
Howell \etal (2001). At this stage the mass transfer ceases and the
secondary star reestablishes the Roche lobe contact, when the period drops
to 2.15~h (Howell \etal 2001, or 2.10~h as found by Knigge 2006). Further
evolution of the system is governed by the angular momentum loss driven by
the emission of gravitational waves.

The periodic brightness changes observed in the quiescence may be caused by
changing visibility of the hot spot (hot part of the disk where a stream of
matter reaches the disk). In systems with $P_{\rm orb}>5$~h the secondary
contributes significant flux to the total flux of the system (Warner 2003)
and the double wave ellipsoidal variations caused by the distorted
secondary can be seen, if the inclination of the system is high enough.
The stars presented in this paper are located in the dense stellar region
and the possibility of blending with other, possibly variable, sources can
not be ruled out. Blending also lowers the amplitude of the outbursts.

During the superoubursts we expect to see superhumps which manifest as a
tooth-shaped light variations with an amplitude of up to 0.5~mag (\eg
Semeniuk \etal 1997). The superhump period, which may be changing, is
typically a few percent longer than the orbital period. This period excess
is greater for systems with longer orbital periods (Stolz and Schoembs
1984).

In the Section~2 we give an overview of photometric data used. The three
following sections describe new DNe. We end with a summary and future
plans.

\Section{Observations}
The main dataset analyzed in this work comes from the OGLE-IV survey which
started in March 2010. This ongoing survey uses 1.3-m Warsaw telescope
located at Las Campanas Observatory, Chile. The observatory is operated by
the Carnegie Institution for Science. The mosaic camera consists of 32
CCD chips $2048\times4102$ pixels each. The total field of view is 1.4
square degrees. Pixel size of $15\mu{\rm m}$ corresponds to 0\zdot\arcs26 in
the sky. Most of the observations are taken with the {\it I}-band filter
with an exposure time of 100~s. Data of some objects acquired between
Julian days 2\,455\,250 and 2\,455\,375 may be affected by technical
problem which manifests itself through a higher scatter of the data or
shifted mean brightness. The main goal of the OGLE-IV survey is to observe
central Galactic bulge fields frequently enough to see planetary
perturbations in the gravitational microlensing events in regular survey
mode \ie without finding them while ongoing and intensifying the
observations. Three fields named BLG501, BLG504 and BLG505 with equatorial
coordinates of the centers $(\alpha, \delta)=(17{:}51{:}53,-29{:}50{:}45)$,
$(17{:}57{:}30,-28{:}00{:}00)$ and $(17{:}57{:}30,-29{:}13{:}50)$,
respectively, are observed with a cadence of 20 minutes\footnote{See the
OGLE webpage {\it http://ogle.astrouw.edu.pl} for the sky coverage and the
cadence of other fields.}. This provides a unique opportunity to study
other transient phenomena like the dwarf novae eruptions with high time
resolution.

\begin{figure}[htb]
\includegraphics[angle=270,width=13cm]{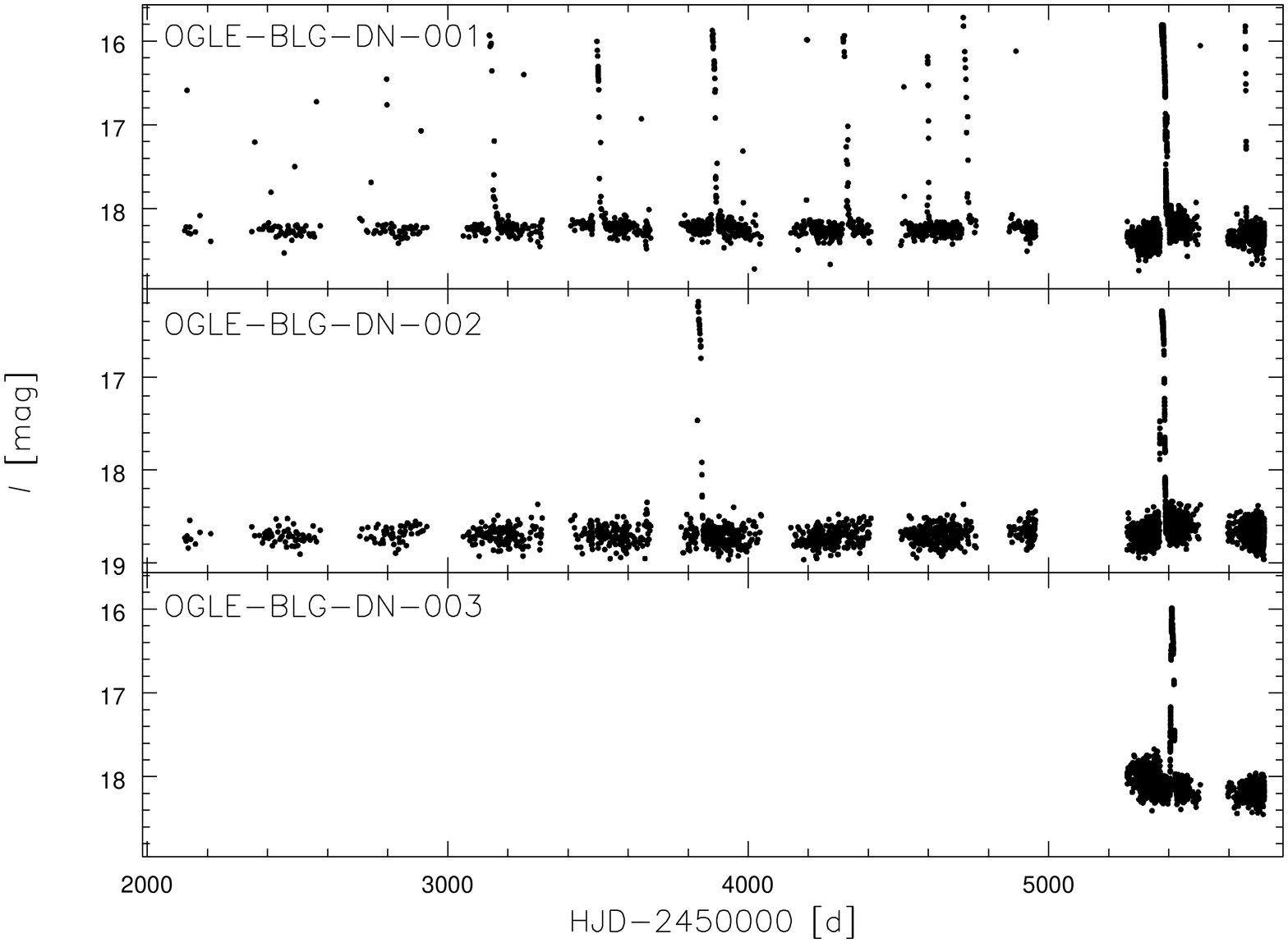}
\FigCap{OGLE-III and OGLE-IV light curves for the analyzed DNe.}
\end{figure}

Central parts of the Galactic bulge were also observed in the previous
phases of the OGLE survey. During the third phase (OGLE-III) the same
telescope was used with the camera containing eight CCD chips.
Observations lasted from 2001 to 2009. Details of the setup and data
reduction were given elsewhere (Udalski 2003, Udalski \etal 2008). The
colors of the dwarf novae change in the {\it a priori} unknown way, thus we
did not apply corrections to instrumental magnitudes described by Udalski
\etal (2008), which may change the zero point of magnitude scale by 0.1~mag.

The photometry was obtained using the Difference Image Analysis (DIA)
me\-thod (Alard and Lupton 1998, Alard 2000, Wo¼niak 2000). We supplemented
the OGLE-IV photometric data with the OGLE-III ones, if available. Whole
analyzed dataset is presented in Fig.~1.

\MakeTableee{lrrr@{}l@{}rrr}{12.5cm}{Basic data of analyzed stars}
{\hline
\multicolumn{1}{c}{ID}\uprule & 
\multicolumn{1}{c}{RA} & 
\multicolumn{1}{c}{Dec}& 
\multicolumn{2}{c}{$I_q$}& 
\multicolumn{1}{c}{$I_{\rm max}$}& 
$N_{\rm O3}$ & 
$N_{\rm O4}$\\
\dorule & 
& 
& 
\multicolumn{2}{c}{[mag]}& 
\multicolumn{1}{c}{[mag]}& 
& \\
\hline
\uprule
OGLE-BLG-DN-001 &  17\uph53\upm10\zdot\ups04 & $-29\arcd21\arcm20\zdot\arcs6$ & $18.62-17.96$ & $^a$ & 15.72 & 1354 & 2487 \\
OGLE-BLG-DN-002 &  17\uph53\upm18\zdot\ups69 & $-29\arcd17\arcm17\zdot\arcs3$ & $18.97-18.34$ & $^a$ & 16.18 & 1354 & 2472 \\
OGLE-BLG-DN-003 &  17\uph56\upm07\zdot\ups27 & $-27\arcd48\arcm47\zdot\arcs5$ & $18.45-17.96$ &      & 15.99 & 0    & 2461 \\
\hline
\noalign{\vskip4pt}
\multicolumn{8}{p{12.5cm}}{$I_q$ is brightness range in quiescence ($^a$ --
blended with other object), $I_{\rm max}$ is maximum recorded brightness,
$N_{\rm O3}$ and $N_{\rm O4}$ are the numbers of measurements in the
OGLE-III and the OGLE-IV data, respectively.}
}
Table~1 gives basic parameters of analyzed stars together with the number
of measurements during the third and the fourth phase of the OGLE survey.
We have checked centroids from the OGLE-III PSF photometry (OGLE-BLG-DN-001
and OGLE-BLG-DN-002) and DIA centroids from OGLE-IV (OGLE-BLG-DN-003). For
OGLE-BLG-DN-001 and OGLE-BLG-DN-002 we have clearly found shift between
apparent centroid of the stars in the quiescence and close to the maximum
brightness. This shows that the two objects are blended and they are
fainter than the brightness in the quiescence ($I_q$) given in Table~1.
Finding charts are presented in Fig.~2.

\begin{figure}[htb]
\begin{center}
\begin{tabular}{@{} l @{~} l @{~} l @{}}
OGLE-BLG-DN-001 & OGLE-BLG-DN-002 & OGLE-BLG-DN-003 \\
\includegraphics[width=4.09cm]{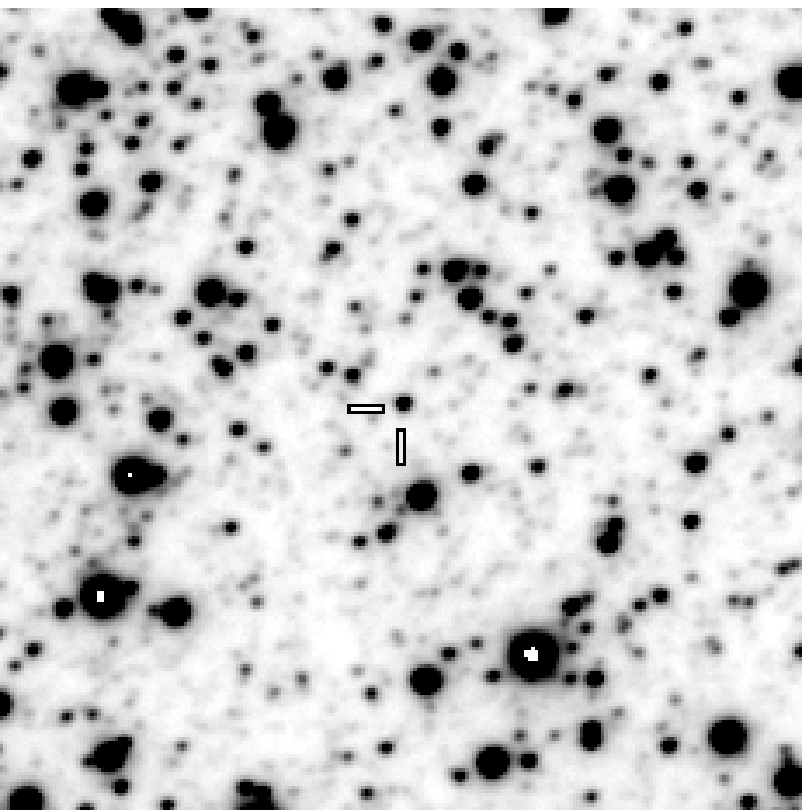} &
\includegraphics[width=4.09cm]{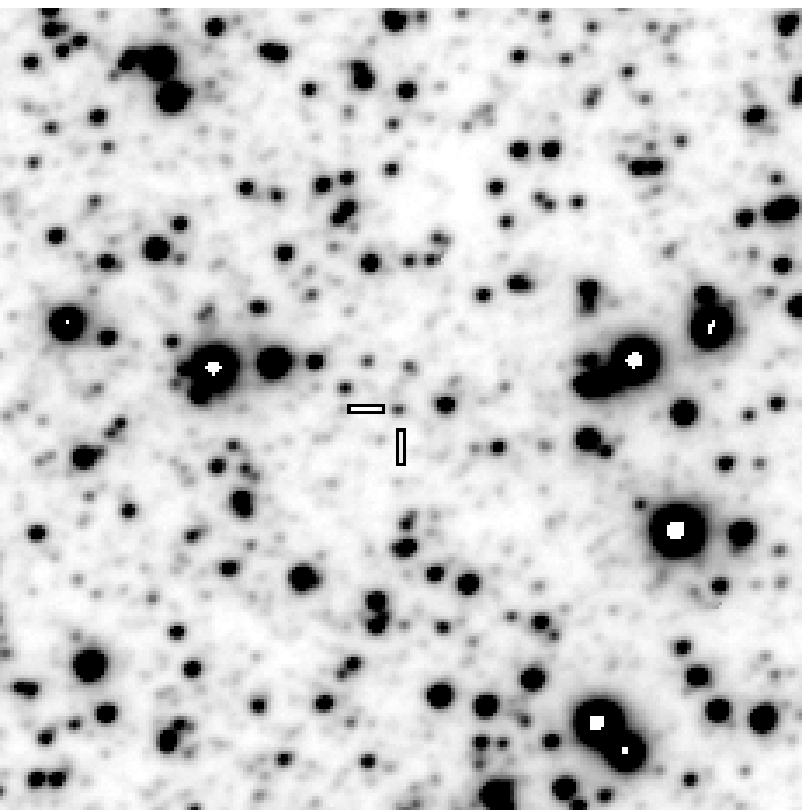} &
\includegraphics[width=4.09cm]{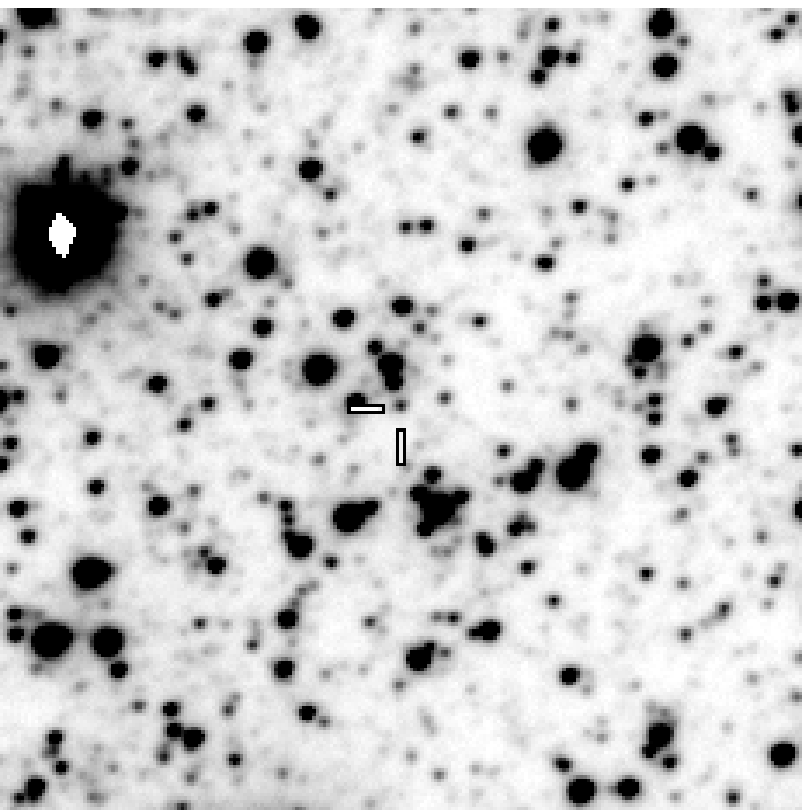}
\end{tabular}
\end{center}
\FigCap{Finding charts of analyzed stars. North is up, East is to the
left. These subframes of {\it I}-band images cover $60\arcs\times60\arcs$.}
\end{figure}

\Section{OGLE-BLG-DN-001}
The whole OGLE light curve of this object is shown in the upper panel of
Fig.~1. For most outbursts we have only one or two observations, thus, to
ensure these are not some kind of an artifacts of the DIA photometry, we
have carefully checked PSF photometry for OGLE-III data (Udalski \etal
2008) and examined the last image of the field taken in 2010 season of
OGLE-IV. In all cases we confirmed that the object was in the bright state
and each point brighter than 17.9~mag belongs to one of the outbursts.

\subsection{Outbursts: Shapes and Frequencies}
Results of timing of outbursts are given in Table~2 which gives ${\rm
HJD'_{max}}$ (hereafter ${\rm HJD'\equiv HJD-2\,450\,000}$) and maximum
recorded brightness $I_{\rm max}$ if it was brighter than 16.5~mag. In
total 23 events were observed. The time interval between definite
superoutbursts is between 354 and 435 days with an exception of 662~d, when
a gap between the OGLE-III and the OGLE-IV occurs. It is possible that the
supercycle is two times shorter and superoutbursts occurred during the
seasonal gaps. In that case the ratio of supercycle length to normal cycle
length would be small compared to other SU~UMa type stars (Warner 2003).
We suppose that also observations done at ${\rm HJD'}=2358$ and 2745 caught
the superoutbursts rather than the normal outbursts. Firstly because the
time difference between them is 387~d and the next superoutburst occurs
392~d later, which fits well to the time intervals of the superoutbursts
given above, and secondly, a few days before and after these observations
\begin{figure}[htb]
\includegraphics[angle=270,width=13cm]{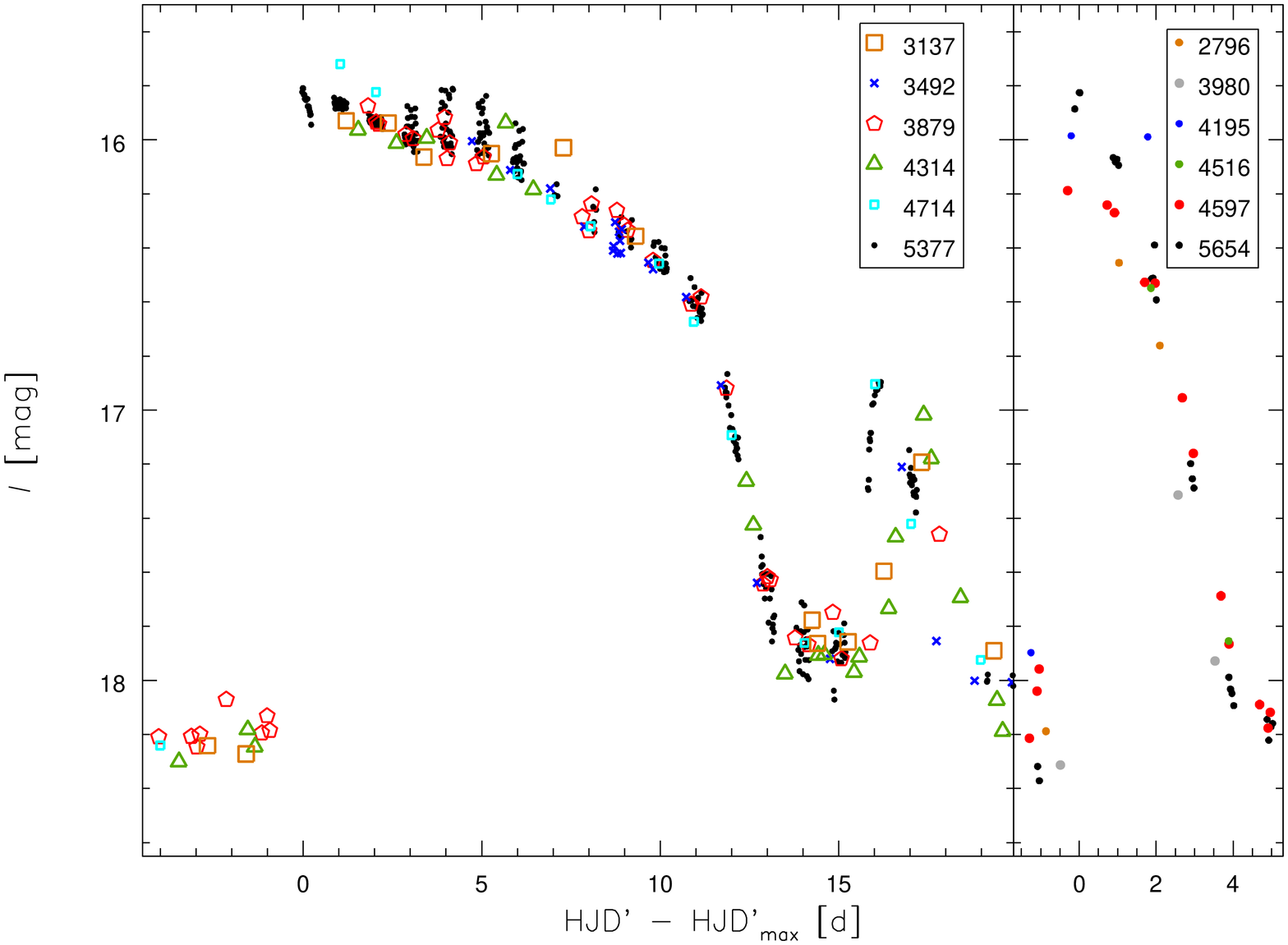}
\FigCap{Aligned profiles of the superoutbursts ({\it left panel}) and the 
normal outbursts ({\it right panel}) of OGLE-BLG-DN-001. Different
outbursts are indicated by different symbols and labeled. Zero on the
horizontal axis corresponds to the time of maximum light.}
\end{figure}
were done no other observation was secured. Fig.~3 shows the profiles of
the superoutbursts (left panel) and the normal outburst (right panel). All
events with more than one observation were shifted to fit the plateau phase
and final decline. The ${\rm HJD}'=2358$ observation fits a superoutburst
profile, if it was done during the brightening of the object before the
plateau and the next observation was done just after the superoutburst
ended. The ${\rm HJD}'=2745$ observation had to be obtained while the
object was fading after the plateau. The point at 11~days earlier was taken
just before the superoutburst started and the whole event was slightly
shorter than the ones shown on the left panel of Fig.~3.
\renewcommand{\arraystretch}{0.95}
\MakeTableee{rrl}{15cm}{Outburst timing of OGLE-BLG-DN-001}
{\hline
\noalign{\vskip3pt}
\multicolumn{1}{c}{${\rm HJD'_{max}}$} & \multicolumn{1}{c}{$I_{\rm max}$} & \multicolumn{1}{c}{comments} \\
\multicolumn{1}{c}{~} & \multicolumn{1}{c}{[mag]} & \multicolumn{1}{c}{~} \\
\hline
\noalign{\vskip3pt}
2133 & & 1 point \\
2358 & & 1 point, probable superoutburst$^a$  \\
2411 & & 1 point \\
2491 & & 1 point \\
2563 & & 1 point \\
2745 & & 1 point, probable superoutburst$^a$\\ 
2796 & 16.46 & 2 points \\
2911 & & 1 point \\
3137 & 15.93 & superoutburst with echo and probable superhumps$^a$ \\
3253 & 16.40 & 1 point \\
3492 & 16.00 & superoutburst with echo \\
3644 & & 1 point \\
3879 & 15.87 & superoutburst with echo \\
3980 & & 2 points \\
4195 & 15.98 & 2 points \\
4314 & 15.96 & superoutburst with echo and probable superhumps$^a$ \\
4516 & & 2 points \\
4597 & 16.19 & \\
4714 & 15.72 & superoutburst with echo \\
4891 & 16.12 & 1 point \\ 
5377 & 15.81 & superoutburst with echo and superhumps \\
5504 & 16.06 & 1 point \\
5654 & 15.82 &  \\ 
\hline
\noalign{\vskip3pt}
\multicolumn{3}{p{7cm}}{$^a$ -- see text for details}
}

The estimation of the time interval between the normal outbursts is more
complicated than for the superoutbursts. We might have missed a few of them
because of a few days long gaps in the OGLE-III data and seasonal breaks
when the Galactic bulge was not observed. The best estimate of a typical
time interval is around 80~days but a better constraint can be given when a
few years long coverage by the OGLE-IV data with at least one measurement
per night will be available.

The left panel of Fig.~3 shows that we have never caught the object during
its brightening to the maximum light of the superoutburst. The shortest
time difference between the last point in quiescence and the first point in
superoutburst is 2.98~d for ${\rm HJD}'=3879$ superoutburst. Typically
object fades by 0.07~mag per day during the plateau phase. The length of
the superoutburst is around 12~days. Each definite superoutburst of the
OGLE-BLG-DN-001 is showing an echo outburst with an amplitude two times
smaller than the normal outbursts. These echo outbursts may have two
distinct profiles one for superoutbursts 3137 and 4314 with all the other
belonging to the second group. Osaki \etal (2001) suggested such echo
outbursts may be caused by changes of viscosity in the cold disk.

The amplitudes of normal outbursts are similar to the ones of the
superoutbursts, which is atypical for the SU~UMa type stars. For most stars
of this type the superoutbursts are brighter by 0.7~mag (Warner 2003). For
OGLE-BLG-DN-001 the normal outbursts are three times shorter than the
superoutbursts.

\subsection{Superhumps}
The ${\rm HJD}'=5377$ superoutburst was observed with a cadence of around
20 minutes. It allows more detail analysis of this event. Fig.~4 presents
the plateau phase with model fitted in the upper panel. The superhumps are
clearly seen between ${\rm HJD}'=5379$ and 5383. These data together with
those from the previous night, when most likely the superhumps started, are
shown in panel {\it b} of Fig.~4 after the fit was subtracted. The second
part of the plateau has much worse coverage and cannot give any firm
conclusions about the existence of the superhumps. Panel {\it c} of Fig.~4
shows the evolution of superhump amplitude. It reaches maximum value a few
days after supermaximum and then slowly decreases.

We determined the times of superhump maximum light and fitted the linear
ephemeris to them:
$${\rm HJD}'=5379.5826(45)+0.10412(25)\cdot E.$$ 

Last panel of Fig.~4 shows the $O-C$ diagram for the derived times of
maximum light (Table~3). The fitted parabola corresponds to the superhump
period $P_{\rm sh}$ changing according to the formula:
$$P_{\rm sh}=0.10680(67)-0.00165(50)\cdot({\rm HJD}'-5379).\eqno(1)$$

A separate analysis of the superhump data was performed using the
multi-harmonic analysis of variance ({\sc mhAoV}) method
(Schwarzenberg-Czerny 1996). Periods were derived for data from each night
and the linear regression resulted in the following ephemeris:
$$P_{\rm sh}=0.10990(52)-0.00256(22)\cdot({\rm HJD}'-5379).\eqno(2)$$

\renewcommand{\arraystretch}{1.05}
\MakeTable{|rr|rr|}{12.5cm}{Superhump maxima of OGLE-BLG-DN-001}
{\hline
E & HJD' & E & HJD' \\
\hline
 0 & 5379.5704 & 19 & 5381.5646 \\
 1 & 5379.6732 & 20 & 5381.6720 \\
 2 & 5379.7884 & 21 & 5381.7767 \\
 9 & 5380.5259 & 29 & 5382.5953 \\ 
10 & 5380.6324 & 30 & 5382.7006 \\ 
11 & 5380.7374 & 31 & 5382.7991 \\ 
12 & 5380.8405 &  &  \\
\hline
}

\begin{figure}[htb]
\centerline{\includegraphics[width=9cm]{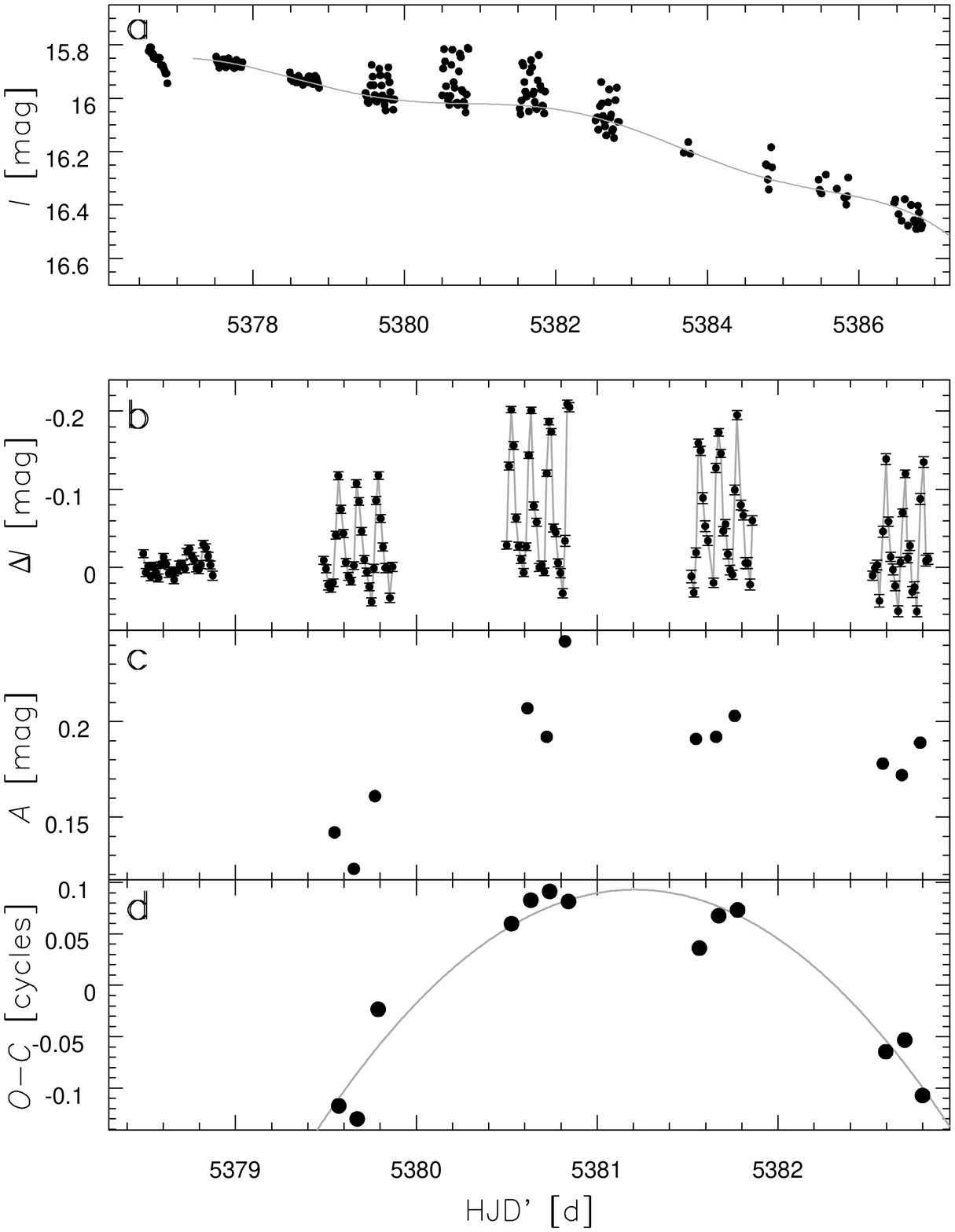}}
\FigCap{Light curve of OGLE-BLG-DN-001 plateau phase ({\it upper panel})
with model fitted to data without superhumps. {\it Lower panels} show a
closeup of the superhump data (consecutive points are connected for
clarity), superhumps amplitude \vs time and $O-C$ diagram for superhumps
maxima, respectively.}
\end{figure}

Results from Eqs.~(1) and (2) altogether give the best estimate of $P_{\rm
sh}$ at the beginning of the superhumps (${\rm HJD}'\approx5379$) equal to
$0.1087~{\rm d}=2.61~{\rm h}$. Even though we have not measured the orbital
period directly, we suggest this object lies inside the period gap. We used
the dependence of the period excess $\epsilon=\left(P_{\rm orb}-P_{\rm
sh}\right)/P_{\rm orb}$ on $P_{\rm orb}$ (Olech \etal 2011), to draw upper
and lower envelope relations and from these relations we found $P_{\rm
orb}$ should be between 0.098 and 0.101~d (2.35 and 2.42~h, respectively).
We note that the period excess should be at least 21\%, if the orbital
period is at the edge of the period gap (2.15~h). This value of the period
excess is not expected for a CV (Olech \etal 2011).

The time derivative of $P_{\rm sh}$ is $-2\cdot10^{-3}$. Two different
methods were used to estimate this value and their results are not the
same. That is because the $O-C$ method uses only brightness maxima and is
sensitive to the changes in their shape, phase etc. The {\sc mhAoV} method
uses whole light curve. Similarly large negative values of the time
derivative of $P_{\rm sh}$ were found only in MN~Dra, NY~Ser and SDSS
J162520.29+120308.7 (Kato \etal 2009, Olech \etal 2011). All these objects
are situated inside the period gap.

We note that there are two observations taken at ${\rm HJD}'\approx 3882.8$
with a time difference of 0.0742~d (\ie 0.71 of $P_{\rm sh}$) and
brightness difference of 0.15~mag. Similarly there are two observations
taken at ${\rm HJD}'\approx 4319.6$ with the differences of 0.2548~d (\ie
2.4 of $P_{\rm sh}$) and 0.19~mag, respectively. We do not expect such
brightness changes to be caused by the decline during the plateau and
suggest that also ${\rm HJD}'=3879$ and 4314 superoutbursts were showing
superhumps.

Using {\sc mhAoV} analysis we tried to search for other periodicities both
after prewhitening plateau data, in quiescence and after subtracting the
echo outburst profile. No sign of a periodic light variations was found.

\Section{OGLE-BLG-DN-002}
This object was announced by the MOA group (Sako \etal 2008) as a candidate
microlensing event with designation MOA-2010-BLG-338. The whole light curve
is shown in the middle panel of Fig.~1 and the close-up of two outbursts
observed by the OGLE-III and the OGLE-IV are shown in Fig.~5.  They are
1543~days apart and, if the same time elapsed after the previous outburst,
it should have taken place during the observing gap between the first two
observing seasons of OGLE-III. Both outbursts show a plateau with
brightness declining 0.05~mag per day and the shape characteristic for
superoutbursts. The length of the second outburst is 2~d longer than the
first one. We suspect that the second outburst might have been triggered by
a precursor. During that time the observations were not conducted because
of the final engineering of the OGLE-IV camera.
\begin{figure}[p]
\centerline{\includegraphics[width=9.5cm,angle=270]{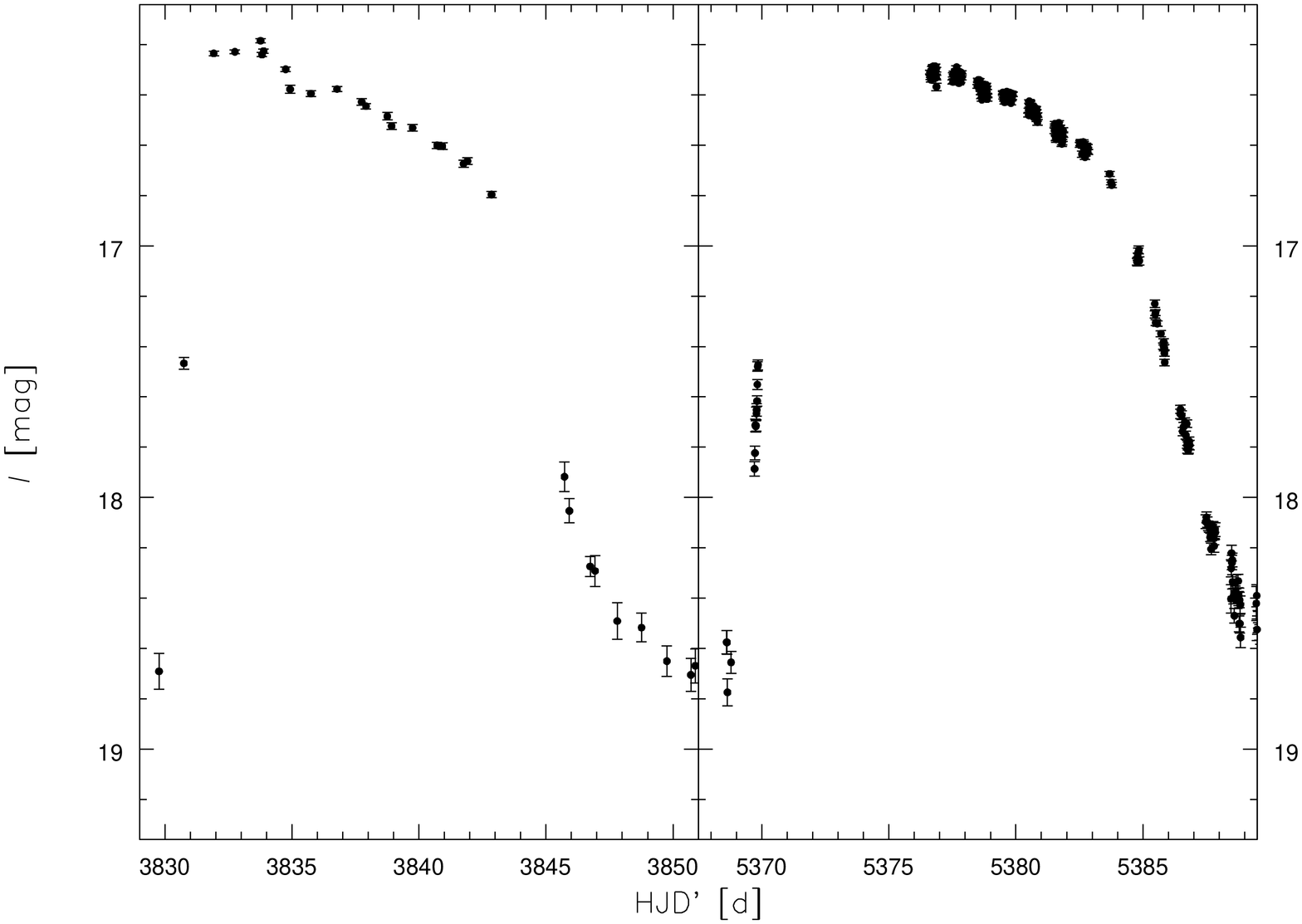}}
\FigCap{Outbursts light curve of OGLE-BLG-DN-002 in the OGLE-III ({\it
left}) and the OGLE-IV ({\it right}). The time span in both panels is the
same.}
\centerline{\includegraphics[width=9.3cm]{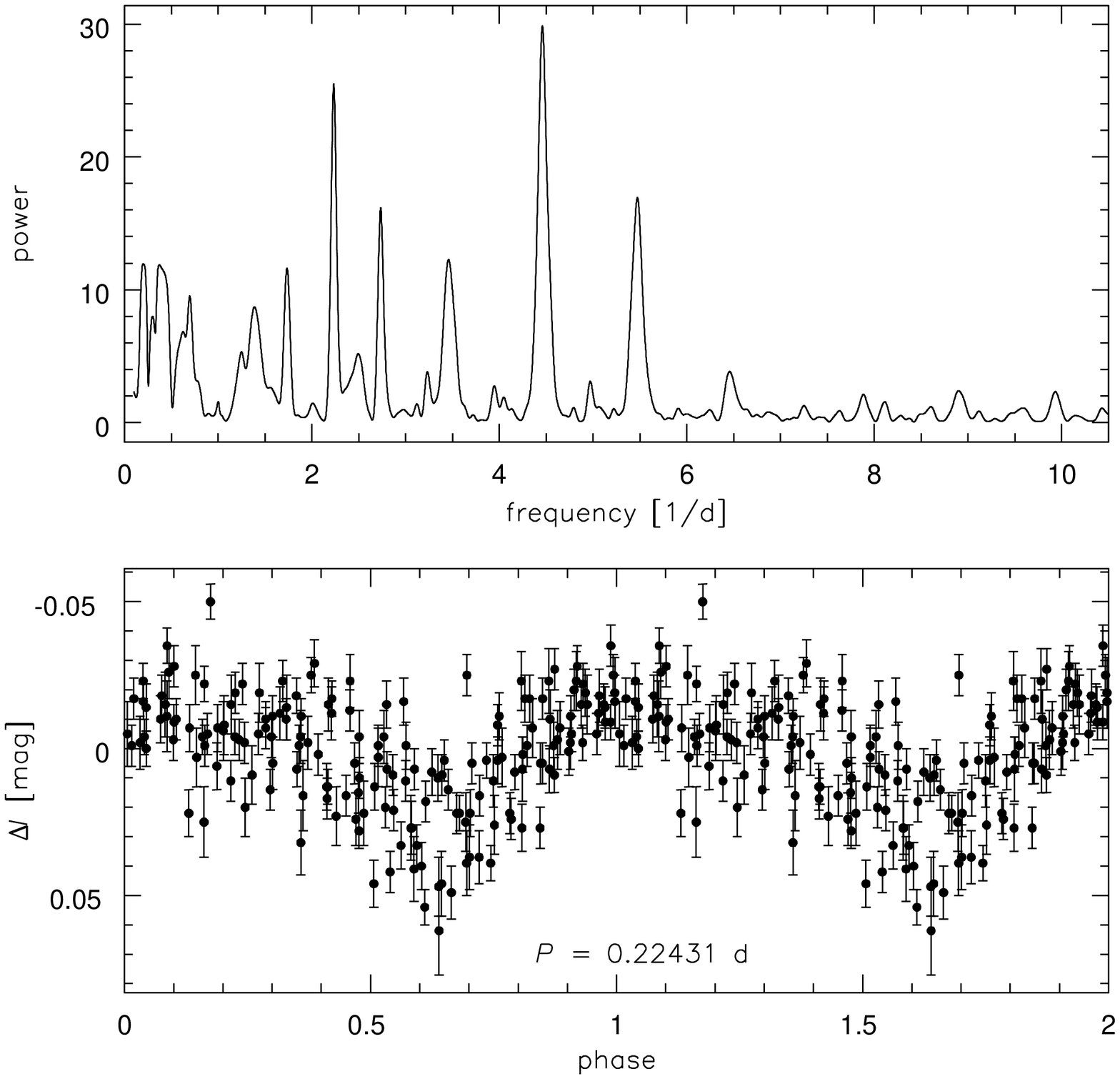}}
\FigCap{Power spectrum of OGLE-BLG-DN-002 plateau data ({\it upper panel}) 
and these data phased with the period 0.22431~d corresponding to the
highest peak in the power spectrum ({\it lower panel}).}
\end{figure}

The plateau data (${\rm HJD}'$ in the range 5376--5383) were analyzed to
search for a possible superhumps. For each night the trend was subtracted
and the {\sc mhAoV} periodogram with two harmonics was calculated. It is
shown in Fig.~6 together with data folded with the period of
$P_1=0.22431(66)$~d corresponding to the highest peak in the power
spectrum. The second highest peak corresponds to the period two times
longer. The profile of the brightness changes is more a tooth-like than a
wave-like and is variable. Contrary to OGLE-BLG-DN-001, for OGLE-BLG-DN-002
it is not clear that the observed light variations were caused by the
superhumps. If we observed the superhumps, then such profiles are more
characteristic for the late part of the plateau (\eg Semeniuk \etal
1997). The length of the $P_1$ does not negate superhump hypothesis. Even
though superhump periods are typically shorter than $P_1$, Retter
\etal (2003) found superhumps with even longer period in TV~Col which is
an intermediate polar (see Warner 2003 for a review). Our analysis of
OGLE-BLG-DN-002 is hampered by a small amplitude of the observed light
variations, which is around 0.08~mag. Reliable determination of times of
maximum light is not possible.

\begin{figure}[htb]
\centerline{\includegraphics[width=9.3cm]{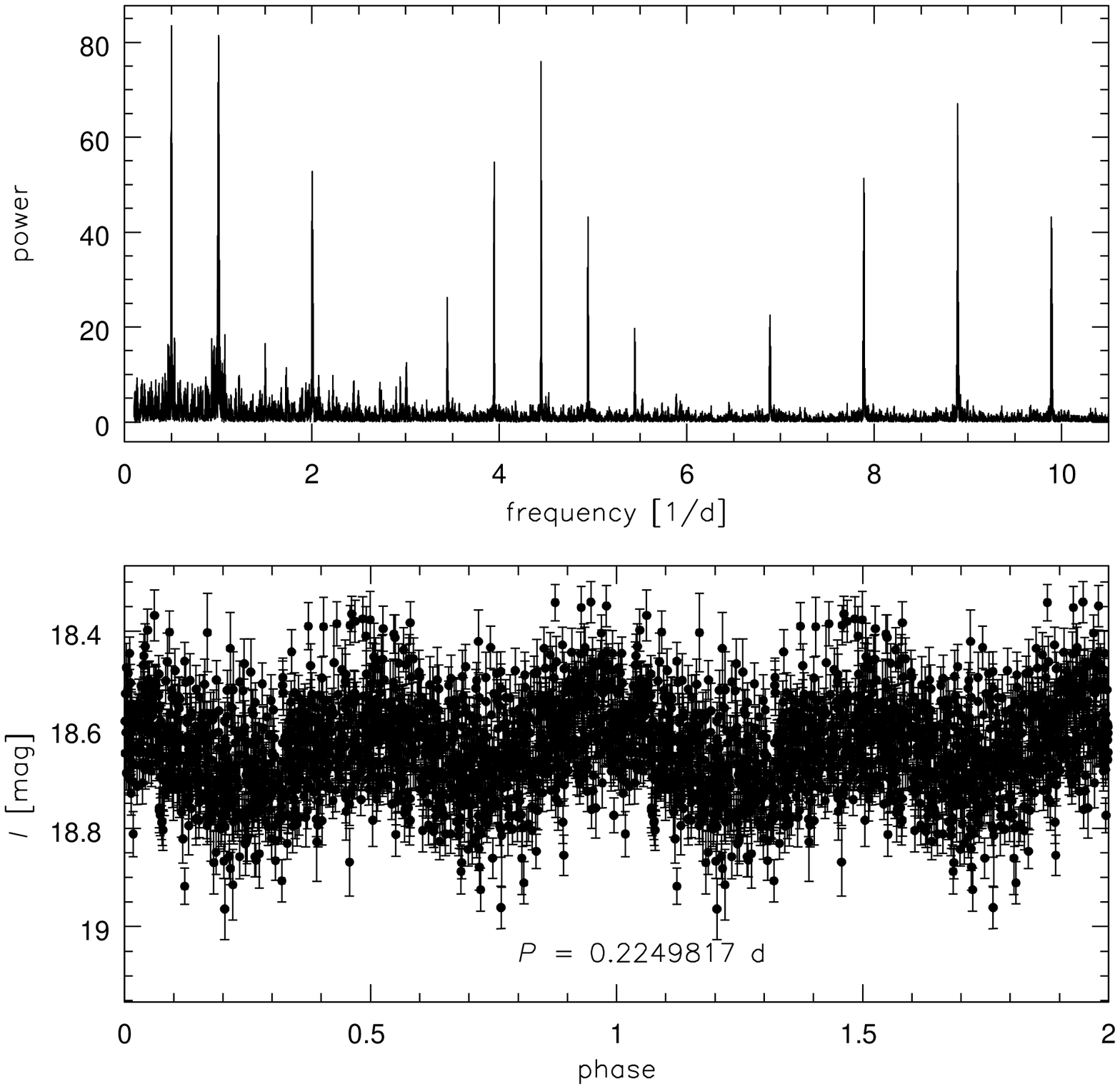}}
\FigCap{Power spectrum of OGLE-BLG-DN-002 quiescence data ({\it upper
panel}) and these data phased with period 0.2249817~d corresponding to the
peak near 4.4~1/d ({\it lower panel}).}
\end{figure}
We have also performed period analysis of the OGLE-IV quiescence data. The
upper panel of Fig.~7 shows the {\sc mhAoV} periodogram calculated with two
harmonics.  The two highest peaks in the periodogram correspond to
$\approx1$~d and $\approx2$~d and they are caused by the observing pattern.
The other peaks are near 4.44~1/d and 8.88~1/d. Also their one and two day
aliases are clearly seen. The lower panel shows data phased with the period
$P_2=0.2249817(35)$~d ($1/P_2\approx4.44$~1/d). A small evidence for an
asymmetry can be seen. The peak near 8.88~1/d corresponds to the quadruple
wave which is very unlikely to be observed.

The period $P_2$ should be equal either $P_{\rm orb}$ or $2P_{\rm orb}$.
During the plateau phase of the superoutburst, we have measured less
precisely variations with a period $P_1$ which is indistinguishable from
$P_2$ within uncertainties. The shape of the light curve suggests
variations with $P_1$ may be superhumps.  Thus, we interpret $P_2$ as being
equal to the $P_{\rm orb}$ and $P_1$, which is less accurately known
because of smaller number of measurements, to be $P_{\rm sh}$.

We prewhitend quiescence and plateau data with periods found and searched
for other periodic variations.  Nothing was found.

\Section{OGLE-BLG-DN-003}
This object was announced by the MOA group as a candidate microlensing
event and designated MOA-2010-BLG-466. Its field was not observed by the
OGLE-III, thus we have only the OGLE-IV data which are shown in the bottom
panel of Fig.~1. Only one outburst was observed and it is shown in detail
in Fig.~8. The upper panel shows the whole outburst and smaller panels
show data from the separate nights. The rising branch lasts longer than
the plateau phase.  It is not obvious, if this is a super- or a normal
outburst.
\begin{figure}[htb]
\vglue-5mm
\centerline{\includegraphics[width=11.5cm]{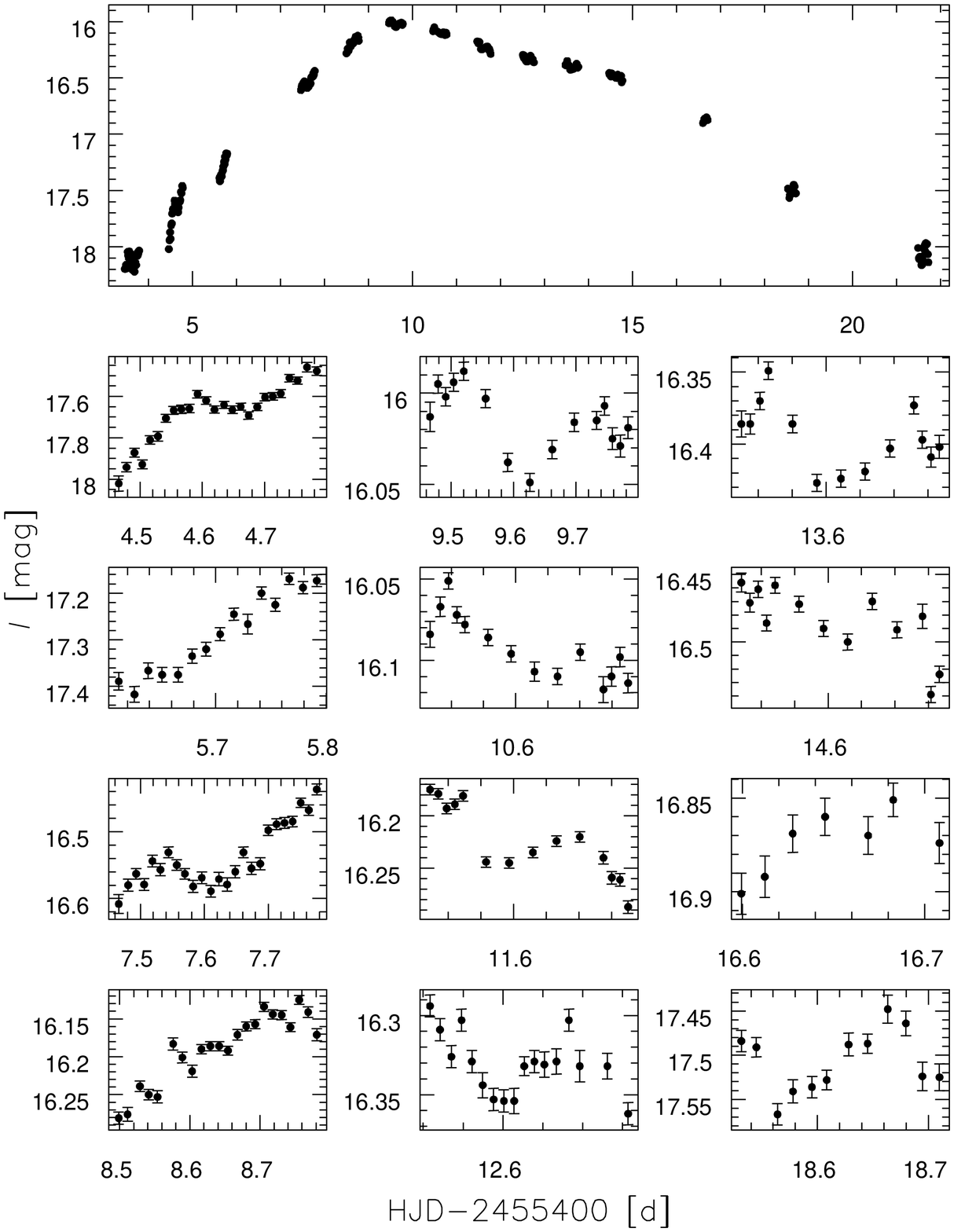}}
\vskip-6pt
\FigCap{Outburst light curve of OGLE-BLG-DN-003. {\it Smaller panels} 
show each outbursts night separately.}
\end{figure}

To search for periodic light changes we subtracted a linear trend from each
night between ${\rm HJD}'=5405$ and 5419 separately. The first night of the
outburst was omitted as it shows very large variations, which may be a
superposition of the changes seen in the quiescence (see below) and
possible superhumps. The {\sc mhAoV} periodograms calculated with one and
two harmonics, as well as data phased with two candidate periods ($P_3=
0.19895(35)$~d and $P_4=0.39803(80)$~d) are shown in Fig.~9. The profiles
of folded light curves do not resemble typical superhump ones.

\begin{figure}[p]
\vglue-3mm
\centerline{\includegraphics[width=9cm]{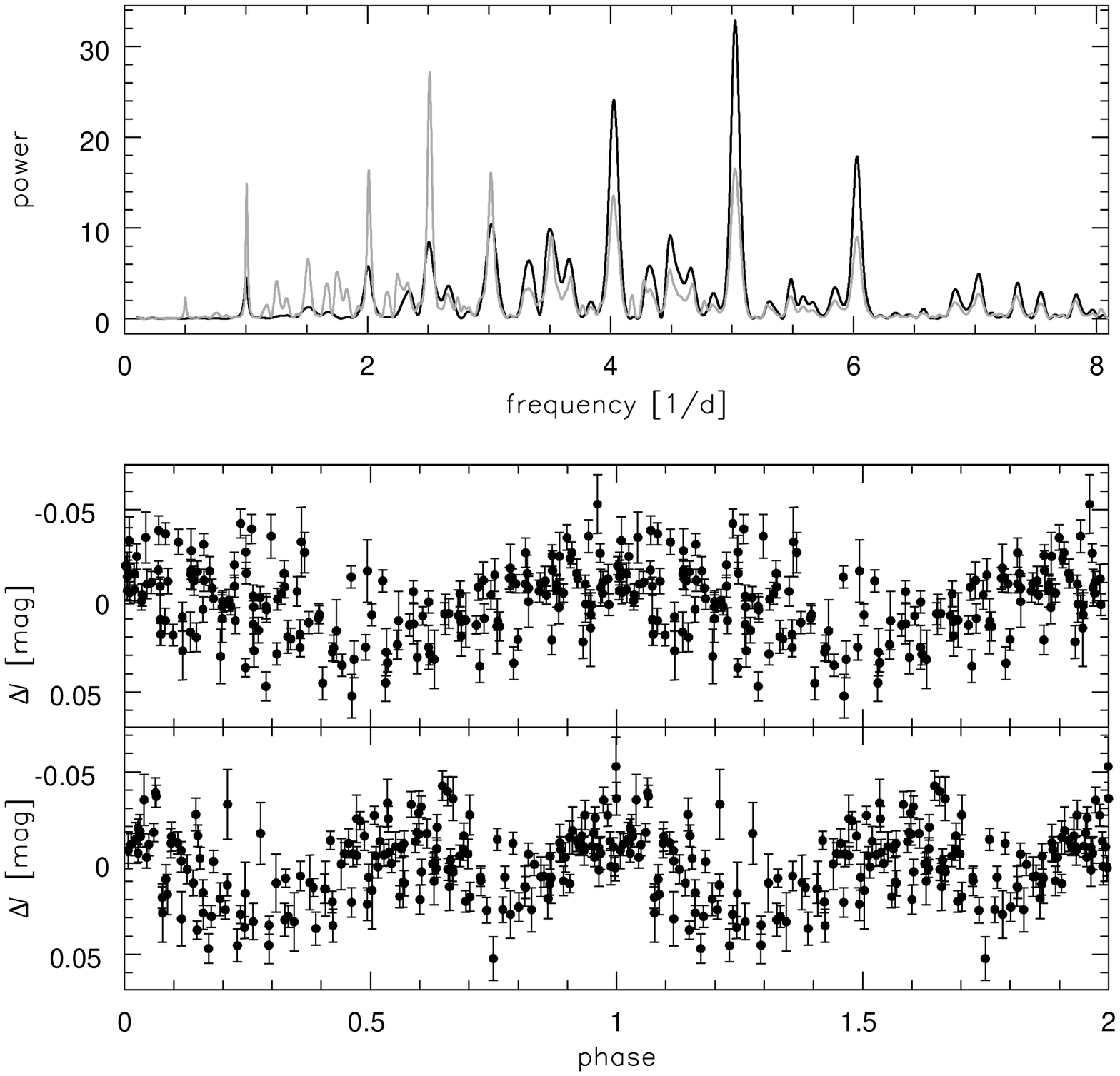}}
\FigCap{Periodogram of plateau data ({\it upper panel}) and the light 
curve phased with two candidate periods: 0.19895~d ({\it middle panel}) and
0.39803~d ({\it lower panel}) for OGLE-BLG-DN-003. The periodogram is shown
for one (black line) and two (gray line) harmonics.}
\vskip5pt
\centerline{\includegraphics[width=9cm]{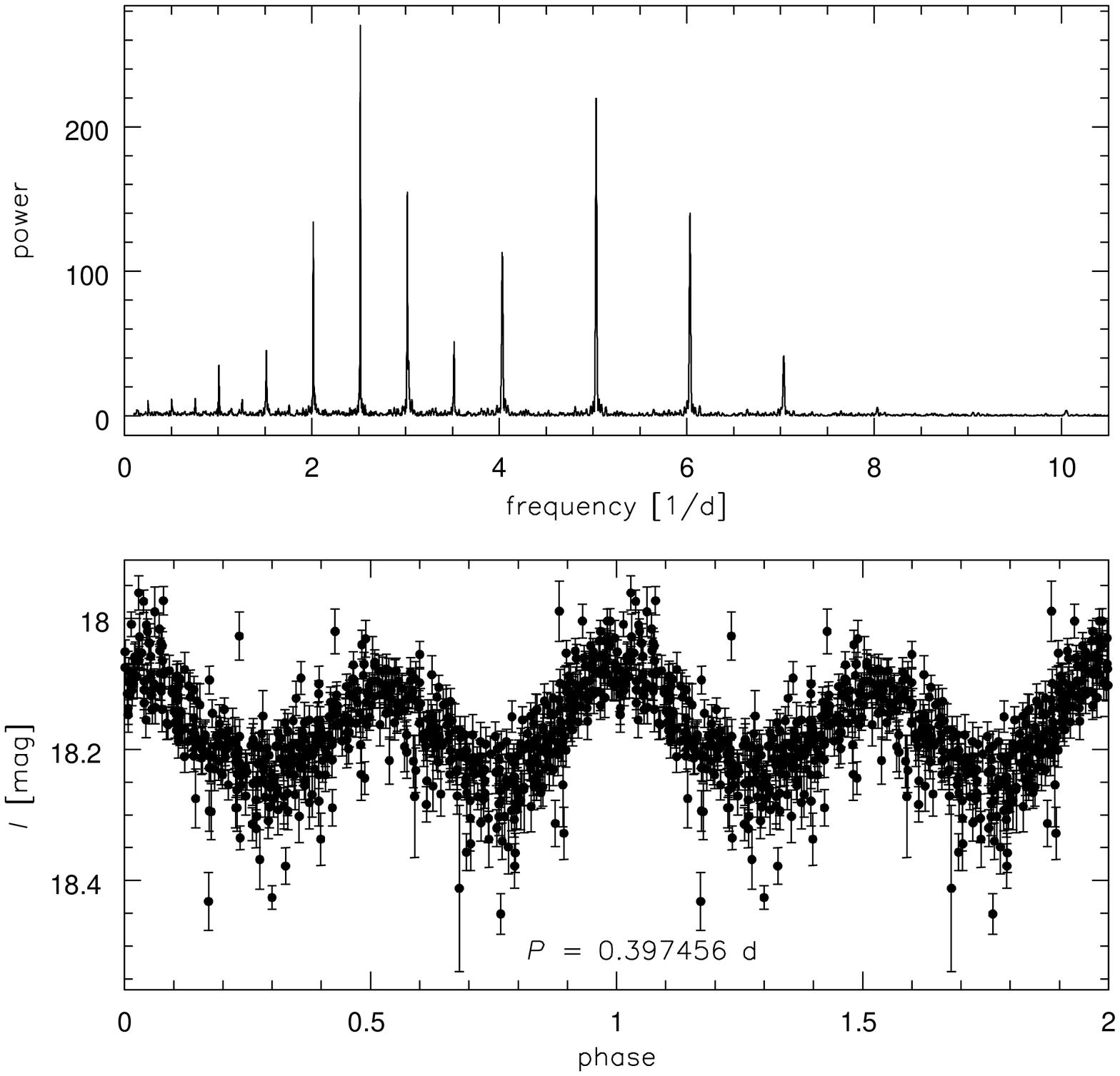}}
\FigCap{Power spectrum of the quiescence data of OGLE-BLG-DN-003 from 2011 
({\it upper panel}) and these data phased with the period of 0.397456(41)~d
corresponding to the highest peak in the power spectrum ({\it lower
panel}).}
\end{figure}

To show brightness variations in the quiescence, we choose 42 days from
2011 season, during which 615 epochs were collected.  Our findings are
applicable to the other part of the quiescence but other light variations
mask them if longer time span of observations is analyzed at once.  The
{\sc mhAoV} periodogram calculated with two harmonics is shown in the upper
panel of Fig.~10. Very high power was found with a main peak corresponding
to period $P_5=0.397456(41)$~d. The analyzed part of the light curve folded
with $P_5$ is shown in the bottom panel. It is clearly asymmetric double
wave. We interpret $P_5$ as an orbital period. It is much longer than the
upper limit of the period gap. Taking into account the odd shape of the
outburst we suspect OGLE-BLG-DN-003 is atypical dwarf nova or possibly an
intermediate polar. We note that the nightly brightness variations seen in
Fig.~8 have amplitude that is consistent with the quiescence variations
dumped by the increasing flux.

\Section{Summary and Future Plans}
We present three new DNe found serendipitously in OGLE data. The most
interesting results were found for OGLE-BLG-DN-001. This SU~UMa subtype DN
shows superoutbursts each $\approx400$~d and normal outbursts with a
recurrence time of around 80~d. The measured $P_{\rm sh}$, when superhumps
emerge, is 2.61~h what is a very strong suggestion that the orbital period
of this object is inside the period gap. The superhump period change rate
is $-2\cdot10^{-3}$ what is a very large negative value for a DN, however
similar to the ones found in other in-the-gap SU UMa type stars.

The orbital periods for OGLE-BLG-DN-002 and OGLE-BLG-DN-003 are most likely
5.40~h and 9.54~h, respectively. These values are well above the upper
boundary of the orbital period gap. According to 7.15 version of the Ritter
and Kolb (2003) catalog, the longest period definite SU UMa type star is TU
Men, which is located in the period gap. The orbital period of
OGLE-BLG-DN-002 is lightly shorter than for TV~Col which shows superhumps
and is classified as an intermediate polar. Vrielmann \etal (2004)
discussed candidate intermediate polars showing superoutbursts. The
outbursts of OGLE-BLG-DN-002 can be classified as superoutbursts. The one
observed for OGLE-BLG-DN-003 is more difficult to be unambiguously
classified. The observational characteristics of OGLE-BLG-DN-002 and
OGLE-BLG-DN-003 do not allow one to convincingly classify these objects in
the currently used scheme.

We have shown that the OGLE-IV photometry acquired with a 20 minute cadence
can be used for investigation of DNe. In the future we plan to publish an
in-depth analysis of the most interesting objects, as well as present a
catalog of such objects in our Galactic bulge fields. The total area
covered by the OGLE-IV survey with a cadence higher than one observation
per night is larger than 70 square degrees, thus, we expect a few hundreds
of DNe to be found. It is also planned to extend the OGLE real time
variable stars monitoring systems (Udalski 2003) to the discovered DNe
allowing real time checking of their current state (outburst, quiescence).

\Acknow{The OGLE project has received funding from the European Research 
Council under the European Community's Seventh Framework Programme
(FP7/2007-2013)/ERC grant agreement No. 246678. RP is supported through
the Polish Science Foundation START program. JS acknowledge support by the
Space Exploration Research Fund of the Ohio State University.}

\end{document}